\documentclass[pdflatex,sn-mathphys-num]{sn-jnl}

\usepackage{graphicx}%
\usepackage{multirow}%
\usepackage{amsmath,amssymb,amsfonts}%
\usepackage{amsthm}%
\usepackage{mathrsfs}%
\usepackage[title]{appendix}%
\usepackage{xcolor}%
\usepackage{textcomp}%
\usepackage{manyfoot}%
\usepackage{booktabs}%
\usepackage{algorithm}%
\usepackage{algorithmicx}%
\usepackage{algpseudocode}%
\usepackage{listings}%
\usepackage{multirow}
\usepackage{amsthm}
\usepackage{mathrsfs}
\usepackage[title]{appendix}
\usepackage{booktabs}
\usepackage{listings}
\usepackage{algorithmicx}
\usepackage{algpseudocode}
\usepackage{xcolor} 
\usepackage{units} 
\usepackage{bm} 
\usepackage{amssymb} 
\usepackage{array}
\usepackage{tikz} 
\usepackage[utf8]{inputenc}
\usepackage{pgfplots} 
\usepackage{multirow}

\newcolumntype{C}[1]{>{\centering\arraybackslash}p{#1}}

\usepackage{xcolor}

\theoremstyle{thmstyleone}%
\theoremstyle{thmstyletwo}%

\theoremstyle{thmstylethree}%

\usepackage[printonlyused]{acronym}
\begin{document}

\newacro{ft} [FT] {Fourier transform}
\newacro{mimo} [MIMO] {multiple-input multiple-output}
\newacro{leo} [LEO] {low Earth orbit}
\newacro{meo} [MEO] {medium Earth orbit}
\newacro{geo} [GEO] {geostationary orbit}
\newacro{ue} [UE] {user equipment}
\newacro{bs} [BS] {base station}
\newacro{isl} [ISL] {inter-satellite link}
\newacro{gnss} [GNSS] {global navigation satellite system}
\newacro{aod} [AoD] {angle-of-departure}
\newacro{aoa} [AoA] {angle-of-arrival}
\newacro{rtk} [RTK] {real-time kinematic}
\newacro{ris} [RIS] {reconfigurable intelligent surface}
\newacro{5g} [5G] {fifth generation}
\newacro{6g} [6G] {sixth generation}
\newacro{ipac} [IPAC] {integrated positioning and communication}
\newacro{psd} [PSD] {power spectral density}
\newacro{los} [LoS] {line-of-sight}
\newacro{nlos} [NLoS] {none-line-of-sight}
\newacro{mcrb} [MCRB] {Misspecified Cramér-Rao Bound}
\newacro{crb} [CRB] {Cramér-Rao Bound}
\newacro{ofdm} [OFDM] {orthogonal frequency-division multiplexing}
\newacro{fdd} [FDD] {frequency division duplexing}
\newacro{tdd} [TDD] {time division duplexing}
\newacro{tle} [TLE] {two-line element}
\newacro{rmse} [RMSE] {root mean square error}
\newacro{pnt} [PNT] {positioning, navigation, and timing}
\newacro{snr} [SNR] {signal-to-noise ratio}
\newacro{tn} [TN] {terrestrial network}
\newacro{ntn} [NTN] {non-terrestrial network}
\newacro{mss} [MSS] {mobile satellite service}
\newacro{soop} [SoOP] {signal of opportunity}
\newacro{3gpp} [3GPP] {3rd Generation Partnership Project}
\newacro{d2d} [D2D] {direct-to-device}
\newacro{d2c} [D2C] {direct-to-cell}
\newacro{soop} [SoOP] {signals of opportunity}
\newacro{psl} [PSL] {positioning service level}
\newacro{esa} [ESA] {European Space Agency}
\newacro{nr} [NR] {New Radio}
\newacro{prs} [PRS] {positioning reference signal}
\newacro{2d} [2D] {two-dimensional}
\newacro{csi} [CSI] {channel state information}
\newacro{foa} [FoA] {formation-of-array}
\newacro{wmmse} [WMMSE] {weighted minimum mean-square error}
\newacro{rms} [RMS] {root-mean-square} 
\newacro{pdop} [PDOP] {position dilution of precision}
\newacro{ttff} [TTFF] {time to first fix}
\newacro{iot} [IoT] {Internet of Things}
\newacro{awgn} [AWGN] {additive white Gaussian noise}
\newacro{nb} [NB] {narrowband}
\newacro{fdd} [FDD] {frequency division duplex}
\newacro{tdd} [TDD] {time division duplex}
\newacro{sinr} [SINR] {signal-to-interference-plus-noise ratio}
\newacro{cdma} [CDMA] {code division multiple access}

\title[Integrated Positioning and Communication via LEO Satellites: A Contemporary Overview]{Integrated Positioning and Communication via LEO Satellites: A Contemporary Overview}


\author[1]{\fnm{Jie} \sur{Ma}}\email{jie.ma@kaust.edu.sa}

\author*[2]{\fnm{Pinjun} \sur{Zheng}}\email{pinjun.zheng@ubc.ca}

\author[3]{\fnm{Xing} \sur{Liu}}\email{xing@ieec.cat}

\author[1]{\fnm{Yuchen} \sur{Zhang}}\email{yuchen.zhang@kaust.edu.sa}

\author[1]{\fnm{Tareq~Y.} \sur{Al-Naffouri}}\email{tareq.alnaffouri@kaust.edu.sa}

\affil[1]{%
\orgdiv{Electrical and Computer Engineering Program, Computer, Electrical and Mathematical Sciences and Engineering (CEMSE)}, %
\orgname{King Abdullah University of Science and Technology (KAUST)}, 
\orgaddress{\city{Thuwal}, \postcode{23955-6900}, \country{Kingdom of Saudi Arabia}}%
}

\affil[2]{%
\orgdiv{School of Engineering}, 
\orgname{The University of British Columbia}, 
\orgaddress{\city{Kelowna}, \state{BC}, \postcode{V1V 1V7}, \country{Canada}}%
}

\affil[3]{%
\orgname{Institut d'Estudis Espacials de Catalunya (IEEC)}, 
\orgaddress{\city{Barcelona}, \country{Spain}}%
}

\abstract{Low Earth orbit (LEO) satellites, as a prominent technology in the sixth generation non-terrestrial network, offer both positioning and communication capabilities. While these two applications have each been extensively studied and have achieved substantial progress in recent years, the potential synergistic benefits of integrating them remain an underexplored yet promising avenue. This article provides a contemporary overview of integrated positioning and communication (IPAC) systems enabled by LEO satellites. Leveraging the distinct characteristics of LEO satellites, we examine how communication systems can enhance positioning accuracy and, conversely, how positioning information can be exploited to improve communication efficiency. A representative case study is included to illustrate enhanced LEO positioning signal acquisition with the assistance of communication-side information. Finally, several key open research challenges for LEO-based IPAC systems are discussed.}


\maketitle
\section{Introduction}
With the rise of applications such as the internet of everything, self-driving vehicles, and spatial intelligence, the demand for accurate positioning and reliable communication continues to grow. As a pivotal technology, \ac{leo} satellites nowadays are playing an important role in the fields of mobile radio positioning and communication in \ac{5g} and beyond~\cite{Liu2021LEO}. 

\Ac{leo} satellites are positioned at orbital altitudes ranging from \unit[300]{km} to \unit[2,000]{km}. Compared with \ac{meo} and \ac{geo} satellites, their lower orbital altitude significantly reduces the signal propagation distance, which results in lower path loss as well as reduced delay. Furthermore, the wide distribution and large number of \ac{leo} satellites enable global coverage and frequent visibility opportunities to ground users. The current \ac{leo} deployments include large communication constellations, such as Starlink, OneWeb, and Telesat, as well as emerging positioning-oriented systems, such as Xona and CentiSpace.
Notably, recent technological advancements have also significantly reduced the costs associated with the satellite manufacturing and launch. The development of smaller, lighter micro-satellites, requiring fewer materials, has facilitated the use of more cost-effective launch vehicles~\cite{ledesma2024trends}. Additionally, the utilization of reusable launch vehicles (RLVs) has further reduced costs and improved the efficiency of space launches. The quick turnaround capability of RLVs also allows for a substantial increase in launch frequency~\cite{SpaceX2021,widiyasmoko2023development}. 

Typically, \ac{leo} satellites can operate in two different paradigms: the ground-based network architecture and the space-based network architecture, as detailed in \ac{3gpp} TR~23.700~\cite{3GPP23700} and illustrated in~Fig.~\ref{fig:workflow}.

In the ground-based network architecture, the satellites use a transparent payload, meaning that satellite transponders are transparent to signals~\cite{3GPP38811}. In this scenario, satellites serve to connect users and \acp{bs} by merely forwarding, filtering, and amplifying signals without decoding/interpreting them. They receive signals from the ground and forward them back to Earth. The \ac{ue} exchanges data through the service link, i.e., the radio link from the \ac{ue} to the satellite. After being forwarded by satellites, the data are transmitted through the feeder link from the satellite network to the terrestrial system. Finally, the data are directed to the end user or the data processing center through the terrestrial data network. 

Unlike the ground-based network, where satellites simply relay signals, the space-based network architecture enables more complex communication tasks. 
The space segment often utilizes regenerative payloads, which not only perform the functions of transparent payloads but also perform signal processing, such as coding/decoding and modulation/demodulation~\cite{3GPP38811}. Additionally, these regenerative payloads can be connected through \acp{isl}, thereby enabling cooperative processing and coordinated transmission across satellites.

With these features, \ac{leo} satellites are emerging as an important platform for both communication and positioning. Recent developments further show that the integration of these two functions is no longer merely conceptual but is gradually moving toward practical deployment~\cite{ESA_Celeste_Introducing, ries2023leo}. Motivated by this transition, this article reviews \ac{leo}-based \ac{ipac} from a system-level perspective, focusing on the mutual enhancement between positioning and communication, the practical constraints of different \ac{leo} architectures, and the remaining technical challenges for future system design.

{\small
\begin{figure*}[t]
  \centering
  \includegraphics[width=1\linewidth]{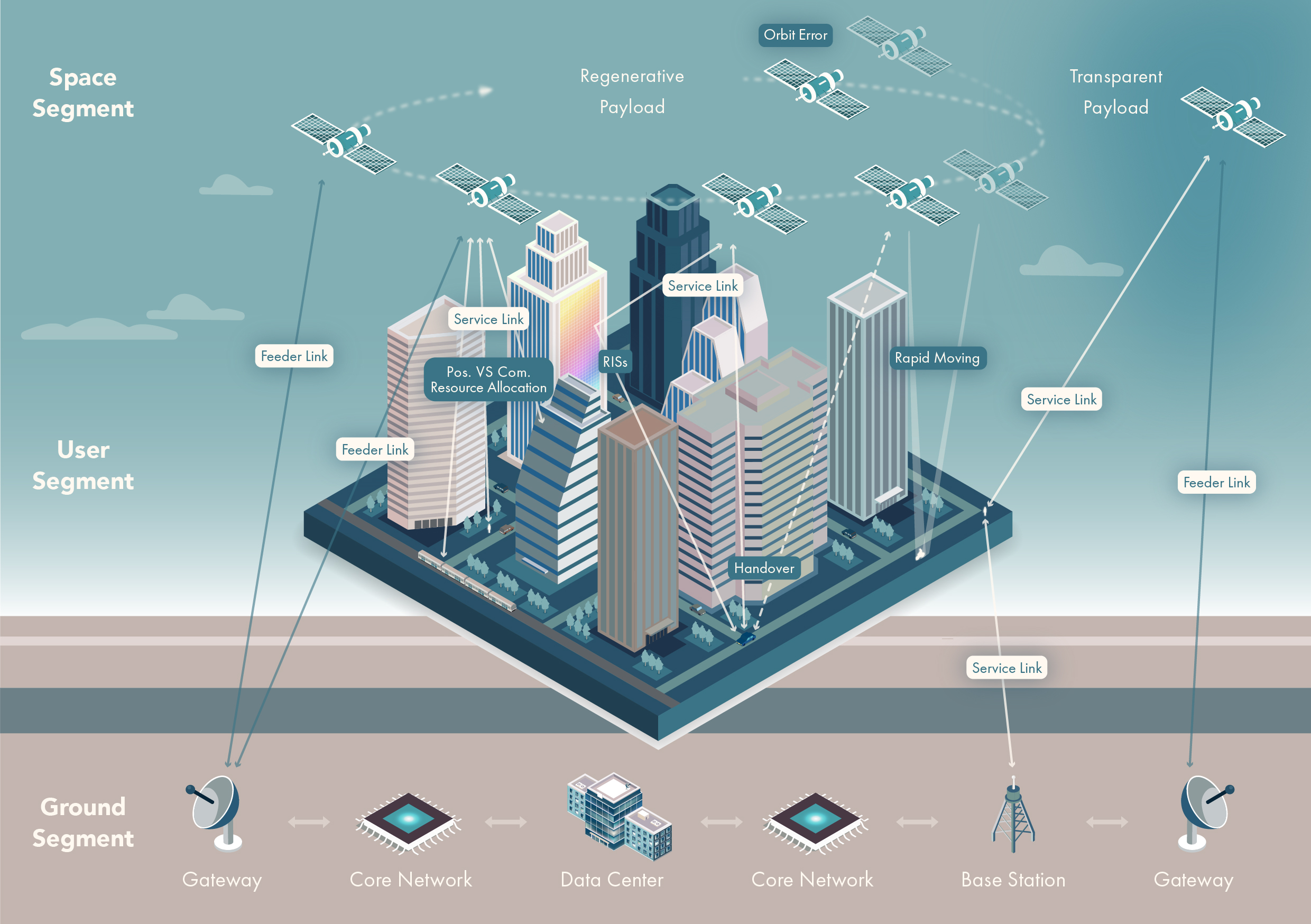}
  \caption{ 
      Structure of LEO satellite systems, divided into the space, user, and ground segments. The space segment of the system provides service and feeder links with various payload types, including regenerative and transparent payloads. The user segment encompasses ground users, infrastructures, and building environments. The ground segment consists of gateways, base stations, data centers, core networks, and other related components.
    }
  \label{fig:workflow}
\end{figure*}

\section{LEO Satellite-Based Positioning and Communication}
This section reviews current and emerging \ac{leo} satellite systems from the perspectives of communication and positioning. We first summarize the major categories of \ac{leo} systems and then discuss their positioning and communication characteristics.

\subsection{LEO System Overview}
The current \ac{leo} satellite constellations are highly heterogeneous in terms of orbital altitude, spectrum allocation, air-interface design, and service types. Based on their system characteristics, as summarized in Table~\ref{table_1}, these constellations can be broadly grouped into three categories: mobile satellite and \ac{d2d}/\ac{ntn} systems, broadband satellite systems, and dedicated or enhanced \ac{pnt} systems.

\begin{sidewaystable*}[t]
\centering
\caption{Representative LEO Satellite Systems}
\label{table_1}
\renewcommand{\arraystretch}{2.5}
\setlength{\tabcolsep}{4pt}
\scriptsize
\begin{tabular}{C{2.2cm}|C{2.8cm}|C{2.0cm}|C{2.3cm}|C{2.2cm}|C{3.3cm}|C{2.3cm}}
\textbf{Type} & \textbf{Name} & \textbf{Orbit} & \textbf{Count} & \textbf{Bands} & \textbf{Waveform / Air interface} & \textbf{PNT / COM} \\
\hline

\multirow{5}{2.2cm}{D2D}
& AST SpaceMobile~\cite{fcc2026ast248,astspacemobile2026network} & 520 / 685 / 690 km & 248 authorized & low-/mid-band & 4G/5G D2D & COM \\
& Starlink D2C~\cite{starlink2026progress} & 550 km & 650+ operational & sub-3 GHz & LTE-based D2C & COM \\
& Lynk~\cite{Lynk_World_Website}& 479--532 km & not publicly disclosed & cellular spectrum & GSM/LTE/5G phone-compatible & COM \\
& Iridium~\cite{iridium2023spares,eoportal2026iridiumnext} & 780 km & 66 operational (+14 spares) & L & TDMA/FDMA-type MSS air interface & COM (+PNT layer) \\
& Globalstar~\cite{globalstar2026technology,globalstar2026who} & 1414 km & 24 operational & L/S & CDMA & COM \\
\hline
\multirow{4}{2.4cm}{Broadband}
& Starlink Gen.~1~\cite{eoportal2026starlink} & 550 km & 4,408 authorized & Ku/Ka & OFDM & COM \\
& Starlink Gen.~2~\cite{fcc2022starlinkgen2,fcc2026starlinkgen2} & 340--614 km & 15,000 authorized & Ku/Ka & OFDM & COM \\
& OneWeb~\cite{eoportal2025oneweb} & 1200 km & 648 planned & Ku/Ka & not publicly specified & COM \\
& Kuiper~\cite{kuiper2020fccauthorization} & 590--630 km & 3,236 authorized & Ka & not publicly specified & COM \\
\hline

\multirow{2}{2.4cm}{\centering\shortstack{Dedicated /\\ Enhanced PNT}}
& Xona Pulsar~\cite{xona2026pulsar} & 1080 km & 258 planned & dual L (X1/X5) & GNSS-compatible dedicated PNT waveform & PNT \\
& ESA Celeste~\cite{esa2026celeste,ESA_Celeste_Constellation} & 500--600 km & 11 planned & L/S/C/UHF & dedicated multi-frequency navigation signals & PNT \\
\end{tabular}

\vspace{1mm}
\begin{minipage}{0.96\textwidth}
\footnotesize
\textit{Note:} PNT denotes positioning, navigation, and timing, and COM denotes communication. The orbit altitudes, constellation sizes, frequency bands, and air-interface descriptions are compiled from publicly available operator information and system documents. All satellite information was accessed in July 2026.
\end{minipage}
\end{sidewaystable*}

\subsubsection{Mobile Satellite and D2D/NTN Systems}
Mobile satellite systems are primarily oriented toward mobile or low-rate services and are typically designed for handheld devices, \ac{iot} devices, and other low-complexity user terminals. They generally operate in lower-frequency cellular or \ac{mss} spectrum, such as the L/S-band or other sub-3 GHz allocations. Traditional \ac{mss} constellations such as Iridium and Globalstar were originally developed for mobile satellite communications using dedicated \ac{mss} spectrum and were designed to support satellite-specific mobile or other low-complexity terminals~\cite{Iridium_Network, Globalstar_Technology}. Recent \ac{d2d}/\ac{ntn} systems, represented by AST SpaceMobile, Starlink~\ac{d2c}, and Lynk, aim to provide direct connectivity to standard mobile phones by maintaining compatibility with cellular air interfaces. In many developments, this is supported through mobile network operators and the use of terrestrial mobile spectrum~\cite{ASTSpaceMobile_Website,Starlink_D2C_Feb2025,Lynk_World_Website}.

\subsubsection{Broadband LEO Systems}
Unlike the mobile satellite and \ac{d2d}/\ac{ntn} systems, broadband \ac{leo} systems, such as Starlink Gen.~1/Gen.~2, OneWeb, and Kuiper, are primarily designed for high-throughput connectivity and mainly operate in the Ku/Ka-band spectrum. Their architectures typically rely on phased-array antennas, multibeam transmission, high-capacity feeder links, and inter-satellite links in some constellations. Rather than targeting direct access by low-complexity handheld terminals, broadband systems are primarily designed for more capable user terminals with directional or electronically steered apertures and generally employ system-specific broadband physical layers~\cite{Starlink_Specifications}. In addition, their service architectures are generally based on scheduled access and beam alignment between the satellite and \ac{ue}.
Therefore, although their large bandwidth and rapidly changing \ac{leo}-\ac{ue} geometry are attractive for opportunistic positioning, the underlying terminal-pointing constraints, communication-oriented waveforms, and scheduled transmissions complicate the extraction of reliable \ac{pnt} observables from their signals.

\subsubsection{Dedicated or Enhanced PNT Systems}
Dedicated or enhanced \ac{pnt} systems treat positioning, navigation, and timing services as their primary objectives. Their signal and system designs are more directly aligned with ranging capability, timing support, and compatibility with existing \ac{gnss} services. This category includes (i)~emerging native \ac{leo}-\ac{pnt} systems, such as Xona Pulsar and Celeste, and (ii)~the enhanced PNT services built on existing satellite infrastructure, such as Iridium PNT. Xona Pulsar is a commercial navigation constellation operating with dual-L-band signals and designed to be compatible with existing L1/L5 \ac{gnss} receivers, while Celeste is designed to demonstrate multi-frequency LEO-PNT signals, including L-, S-, C-band, and UHF components~\cite{xona_technology, ESA_Celeste_Resilience}.

\subsection{LEO Satellite-Based Positioning}
In recent years, \ac{leo}-based positioning, as part of the broader framework of \ac{leo} satellite \ac{pnt}, has attracted significant attention for its potential to complement \ac{gnss} or serve as a reliable backup in the event of \ac{gnss} system failure~\cite{allahvirdi2025doppler,foreman2025comparative}.
Positioning with \ac{leo} satellites has distinct characteristics introduced by the lower orbital regime, some of which may be advantageous relative to systems based on higher orbits such as \ac{meo} and \ac{geo} satellites.
For example, the fast-moving nature of \ac{leo} satellites not only enables the use of stronger Doppler shifts~\cite{shi2023revisiting,zhao2023analysis} but also provides faster geometry variation and stronger temporal diversity, which improve observability and help accelerate positioning convergence~\cite{xu2025joint}. The lower altitude of \ac{leo} satellites can also facilitate integrated positioning and tracking between \acp{tn} and \acp{ntn} by reducing propagation latency and improving the efficiency of cross-segment cooperation~\cite{11049853,saleh2025integrated,ettefagh2025}. 
Moreover, the large size of emerging \ac{leo} constellations may provide increased observable redundancy, although it also introduces more frequent handovers and higher network-management complexity. Such redundancy may also improve positioning robustness against interference, jamming, and spoofing attacks~\cite{ries2023leo}.

However, the characteristics that make \ac{leo} satellites attractive for positioning also introduce several practical constraints. Due to the lower orbital altitude, LEO satellites have smaller instantaneous coverage footprints than \ac{meo}/\ac{geo} satellites, while their rapid orbital motion leads to shorter visibility durations for ground users. Since standalone three-dimensional position estimation typically requires at least four independent satellite measurements, these coverage and visibility constraints mean that continuous \ac{leo}-based positioning generally requires a much larger constellation to ensure sufficient simultaneous satellite visibility. The benefit of the lower orbital altitude also depends on the coverage geometry and antenna design. 
Compared with \ac{meo} satellites, \ac{leo} satellites require their transmit antennas to support broad angular coverage, especially toward the low-elevation footprint edge. For a fixed antenna, this may require a broad sector-shaped radiation pattern, while for a steerable antenna array, it requires a sufficiently wide beam-scanning range.

From an implementation perspective, \ac{leo}-based positioning can be realized through several strategies~\cite{10542356,soualle2024new}. The first strategy relies on dedicated \ac{leo}-\ac{pnt} satellites that broadcast navigation signals specifically designed for ranging and timing. The second strategy enhances or jointly designs communication-capable \ac{leo} satellites with \ac{pnt} capabilities, either by integrating an additional \ac{pnt}-specific payload~\cite{prol2024leo,prol2025leo} or by embedding \ac{pnt} functionalities directly into the communication payload. Such \ac{ipac}-oriented designs have been investigated for \ac{d2d}/\ac{ntn}-oriented \ac{leo} systems~\cite{de2026beyond1,de2026beyond,de2025integrated}, and related concepts have also been studied for broadband satellite systems~\cite{Liao2023}. The third strategy exploits signals from non-dedicated communication satellites as signals of opportunity, where existing broadband or mobile-satellite signals are processed to extract positioning observables without dedicated navigation signals~\cite{kassas2025exploiting,desai2026snapshot}. 

\begin{table*}[t]
\centering
\caption{Representative Positioning and Timing Requirements in 5G/6G-related Standards~\cite{3gppts122261,3gppts22104,imt2030}.}
\label{table_2}
\renewcommand{\arraystretch}{2.8}
\setlength{\tabcolsep}{3.5pt}
\footnotesize
\begin{tabular}{C{3.3cm}| C{3cm}| C{2.5cm}| C{2.5cm}}
\hline
\textbf{Scenario} &
\textbf{Positioning Requirement} &
\textbf{Timing / Sync Requirement} &
\textbf{Source}
 \\
\hline
Outdoor and urban positioning scenarios; 
Positioning Service Level 5 &
0.3 m horizontal, 2 m vertical at 95\% confidence &
-- &
3GPP TS 22.261 \\
\hline
High-availability and low-latency positioning;
Positioning Service Level 6  &
0.3 m horizontal, 2 m vertical at 95\% confidence, with latency below 10 ms &
-- &
3GPP TS 22.261 \\
\hline
UE-to-UE relative positioning within a short range; Positioning Service Level 7   &
0.2 m horizontal and vertical relative positioning at 95\% confidence &
-- &
3GPP TS 22.261 \\
\hline
Global clock-domain synchronicity for industrial automation &
-- &
$\leq 900$ ns end-to-end synchronization error &
3GPP TS 22.104\\
\hline
Long-term 6G positioning target &
1--10 cm positioning accuracy &
-- &
ITU-R M.2160-0, IMT-2030 \\
\hline
\end{tabular}
\end{table*}

Across these implementation strategies, practical \ac{leo}-based positioning systems must be evaluated with respect to the positioning and timing requirements of emerging services. As summarized in Table~\ref{table_2}, current \ac{5g} standards and emerging \ac{6g} visions envision sub-meter or even centimeter-level positioning, along with sub-microsecond to nanosecond timing. For example, \ac{3gpp} TS~122.261 Release-17 defines \ac{5g} \acp{psl} with horizontal accuracy ranging from \unit[10]{m} to \unit[0.3]{m}~\cite{3gppts122261}, and \ac{3gpp}~TS~22.104 for industrial automation requires end-to-end clock synchronization error below \unit[900]{ns}~\cite{3gppts22104}. IMT-2030 further specifies positioning accuracy requirements on the order of \unit[1--10]{cm}~\cite{imt2030}. 

Recent studies have reported promising positioning results under \ac{leo}-based positioning architectures. For dedicated \ac{leo}-\ac{pnt}, centimeter-level orbit determination, sub-nanosecond synchronization, and sub-meter-level positioning have been reported using onboard synchronization and network-aided pseudorange frameworks~\cite{MONTENBRUCK20257144,komodromos2025network}. 
Studies on \ac{d2d}/\ac{ntn} positioning have demonstrated standalone \ac{ue} positioning accuracy at the tens-of-meters level~\cite{grec2025direct,edjekouane2025user,de2026beyond}. 
For \ac{soop}-based positioning, code-phase measurements have been extracted from live \ac{5g} \ac{nr}-\ac{ntn} Ka-band signals~\cite{lapin2024code}. In addition, meter-level \ac{2d} positioning has been demonstrated using non-dedicated Starlink and OneWeb Ku-band downlink signals~\cite{kozhaya2024opportunistic}. 
These studies show the potential of \ac{leo}-based positioning, but also highlight its strong dependence on satellite ephemerides, clock information, synchronization, signal models, beam-operation models, and spectrum planning, which will be further discussed in later sections.

\subsection{LEO Satellite-Based Communication}\label{subsec_LEO_based_com}
Compared with terrestrial wireless communication, \ac{leo} satellite communication is characterized by larger free-space path loss, longer propagation delay, larger and more rapidly varying Doppler shifts, and faster channel and network dynamics. These characteristics affect both communication performance and the extraction of positioning observables from communication signals. We therefore discuss the main channel and mobility-related characteristics of \ac{leo} satellite communication, including path loss and fading, propagation delay, and Doppler effects.

\begin{sidewaystable*}[p]
\centering
\caption{The Large-Scale Path Loss of the Satellite-Terrestrial Wireless Channel~\cite{3GPP38811,ITU2019Attenuation}}
\renewcommand{\arraystretch}{1.4}
\begin{tabular}{m{0.17\textwidth}<{\centering}|m{0.8\textwidth}<{\centering}}
\textbf{Item} & \textbf{Features} \\
\hline
 Free space path loss & {\scriptsize Depends on the separation distance and signal frequency.} \\
\hline
Shadow fading loss  & \vspace{0.5em}{\scriptsize\begin{itemize}\setlength\itemsep{0.6em} 
\item Depends on the geometric elevation angle, building intensity, and signal frequency. 
\item Usually modeled as a normal distribution.
\end{itemize}} \\
\hline
 Clutter loss & \vspace{0.5em} {\scriptsize\begin{itemize}\setlength\itemsep{0.6em}
	\item Depends on the geometric elevation angle and signal frequency. 
	\item In the LoS condition, this term is negligible and should be set to \unit[0]{dB}.
\end{itemize}}\\
\hline
 Attenuation due to atmospheric absorption  & \vspace{0.5em} {\scriptsize\begin{itemize}\setlength\itemsep{0.6em} 
	\item Depends on elevation angle, altitude above sea level, water vapor density (absolute humidity), and signal frequency.
	\item It can generally be disregarded at frequencies below 10 GHz. However, for elevation angles less than 10 degrees, it is advisable to account for it at any frequency above 1 GHz.
\end{itemize}}  \\
\hline
 Attenuation due to either ionospheric or tropospheric scintillation & \vspace{0.5em}{\scriptsize\begin{itemize}\setlength\itemsep{0.6em}
	\item Corresponds to rapid fluctuations of the received signal amplitude and phase.
	\item Scintillations depend on location, time-of-day, season, solar and geomagnetic activity.
	\item Ionospheric scintillation shall only be considered for frequencies below 6 GHz, while tropospheric scintillation shall only be considered for frequencies above 6 GHz.
\end{itemize}} \\
\hline
 Building penetration loss  & \vspace{0.5em} {\scriptsize\begin{itemize}\setlength\itemsep{0.6em}
	\item It varies significantly depending on the location within the building and the construction details of the building.
	\item Modern, thermally efficient buildings with materials such as metalized glass and foil-backed panels generally exhibit significantly higher penetration loss than `traditional' buildings.
\end{itemize}} \\
\end{tabular}
\label{table_3}
\end{sidewaystable*}

The attenuation behavior of \ac{leo} satellite channels is commonly characterized by both large-scale path loss and small-scale fading. According to \ac{3gpp} TR~38.811~\cite{3GPP38811}, the large-scale path loss of satellite-terrestrial channels comprises several components: free-space path loss, shadow fading loss, clutter loss, atmospheric absorption, ionospheric/tropospheric scintillation, and potential building penetration loss. A summary of these attenuation effects and their key characteristics is listed in Table~\ref{table_3}. In addition to large-scale path loss, the small-scale channel behavior, i.e., the short-term fluctuations of the received signal amplitude and phase, must also be considered~\cite{goldsmith2005wireless}. For mobile satellite links with handheld or other low-gain terminals, the channel is commonly modeled using land mobile-satellite statistical models that capture mixed propagation conditions, including \ac{los}, shadowed, and blocked states~\cite{ITU2019P681}. For broadband satellite links with directive or electronically steerable user terminals, the narrow receive beam can substantially suppress multipath components. Therefore, the channel is often modeled as an \ac{awgn} channel after accounting for path loss, propagation delay, Doppler shift, and antenna pointing~\cite{3GPP38811}.

The long propagation distance in satellite channels results in substantial propagation delays, which increase communication latency. Additionally, due to the lower orbital altitude and faster apparent motion of \ac{leo} satellites, the propagation distance varies more rapidly over time and can differ significantly among users within the satellite footprint. Such characteristics can limit the support of delay-sensitive services, while also increasing the difficulty of synchronization and delay-aware scheduling. To quantify the resulting delay performance, delay-related metrics, such as the delay satisfaction ratio and delay variance, are often adopted as key performance indicators for evaluating \ac{leo} satellite networks~\cite{10285043}.

Another unique feature of \ac{leo} satellite communication is the strong Doppler frequency shift due to the high orbital velocity of \ac{leo} satellites. Typically, the maximum Doppler shift in satellite signals increases with higher operating frequencies, faster satellite velocities associated with lower orbital altitudes, and smaller elevation angles~\cite{Shi2024OTFS}. For instance, a \ac{leo} satellite operating at \unit[30]{GHz} with an orbital altitude of \unit[800]{km} exhibits a maximum Doppler frequency shift of approximately \unit[660]{kHz}. Such large Doppler shifts are generally harmful to communication since they introduce large carrier-frequency offsets and rapid phase variations, and therefore require accurate frequency tracking and compensation to maintain signal integrity. Furthermore, the wireless channel coherence time is approximately inversely proportional to the channel's Doppler spread~\cite{goldsmith2005wireless}. Therefore, residual Doppler spread or rapid Doppler variation can shorten the effective coherence time and increase the burden of channel tracking and frequency synchronization.

It is worth noting that the above communication-oriented timing and Doppler pre-compensation mechanisms also have important implications for \ac{soop} positioning. To reduce handover rates, LEO beams can be operated as Earth-fixed or moving cells. Under such beam-based operation, timing and Doppler pre-compensation are often applied with respect to the beam center or another reference point within the service area. For an Earth-fixed beam, this reference point remains approximately fixed on the ground while the satellite moves overhead, so the corresponding delay and Doppler compensation become time-varying. Consequently, the delay and Doppler extracted from communication signals may include time-varying beam-dependent biases and may not directly correspond to the real satellite-\ac{ue} geometry. This creates a fundamental constraint for \ac{soop} positioning. A passive receiver may not have access to the beam reference point, beam scheduling, or pre-compensation parameters required to remove these biases. Therefore, unless such beam-operation and pre-compensation information is modeled and available, communication-derived delay and Doppler observables cannot be directly used as geometric measurements for \ac{soop} positioning.

\section{Why Should LEO Positioning and Communication Be Integrated?}
While \ac{leo} satellites provide promising positioning and communication capabilities, these two functions are often discussed separately. However, their integration can provide mutual benefits when the underlying signal, network, and architectural constraints are properly considered. In this section, we examine \ac{leo}-based \ac{ipac} from both directions: how positioning information can support communication, and how communication infrastructure and side information can enhance positioning.
\subsection{The Role of Positioning in Enhancing Communication}
In this subsection, we discuss how position information can support \ac{leo} communication through several representative applications.

\subsubsection{Network Operation and Regulatory Support}
Reliable \ac{ue} position information is increasingly important for the operation of \ac{ntn} communication systems. For instance, position-aware beamforming, beam selection, handover management, and core-network selection can all benefit from accurate \ac{ue} position information. In particular, since \ac{ntn} cells can cover a large geographic area and may span different national or regulatory regions, \ac{ue} position information is needed for service authorization, emergency services, and public warning services~\cite{ETSI_TR122926}. However, \ac{3gpp}~TR~38.882 notes that relying only on \ac{ue}-reported positions (e.g., obtained via \ac{gnss}) is not reliable, since the reported position may be affected by interference, spoofing, or intentional tampering~\cite{3GPP38882}. This uncertainty can limit the reliability of position-dependent communication procedures and regulatory decisions. Therefore, enabling LEO satellites to position users independently can enhance \ac{leo} communication systems by improving the trustworthiness of position-dependent network operation.

\subsubsection{Position Information-Assisted Beamforming}
In \ac{leo} communication systems, beamforming cannot simply follow the \ac{csi}-acquisition procedures commonly used in terrestrial networks. In terrestrial networks, downlink beamforming is commonly supported by accurate channel knowledge, obtained either through uplink-downlink reciprocity in \ac{tdd} systems or through downlink channel feedback in \ac{fdd} systems. However, \ac{leo} satellite channels are characterized by short coherence times and long propagation delays, which make real-time channel estimation and user-specific beamforming more challenging. In current satellite communication systems, beamforming is therefore often realized through fixed beam patterns or geometry-based beam steering, rather than relying on instantaneous user-specific \ac{csi}~\cite{Neinavaie2022StarlinkBeamforming}.

Under these constraints, position information can provide useful geometric side information for \ac{leo} beamforming. Given the positions of the users and the motion of the satellites, directional beamformers can be designed predictively and dynamically toward the \ac{los} direction~\cite{zack2025pace}. Considering the \ac{los}-dominant propagation and predictable satellite motion, geometry-aided beamforming can reduce the dependence on full instantaneous \ac{csi} while maintaining directional link quality~\cite{11400683}.

\subsubsection{Position-Aided Fast Timing Advance Update}
In \ac{5g} \ac{nr} systems, one process that is significantly impacted by the large propagation delay is the random access procedure. To reduce timing misalignment interference, \ac{5g} \ac{nr} employs timing advance for uplink synchronization~\cite{3GPP38213}. Specifically, a timing advance offset is applied to each uplink transmission to compensate for propagation delay differences, so that the received uplink signals remain aligned within the cyclic-prefix window at the base station~\cite{3GPP38213}. Considering the rapid drift and high variance of the propagation delays in \ac{leo} channels, the timing advance cannot be treated as a static parameter. To address this issue, \ac{3gpp}~TR~38.811 has already highlighted the use of \ac{ue} position information and satellite ephemeris~\cite{3GPP38811}. Given \ac{ue} position information, the delays can be computed and predicted, thus enabling fast timing advance update. Initial studies have demonstrated that by effectively utilizing UE position information, a \ac{5g} \ac{nr}-adapted timing advance estimation scheme can be implemented~\cite{Wang2021Location}. 

\subsubsection{Other Benefits}
Beyond regulatory support, beamforming, and timing-advance updates, position information can also improve other aspects of communication. For example, acquiring an accurate UE position can be critical for Doppler compensation of \ac{leo} satellite signals. In addition, accurate satellite and \ac{ue} position information is crucial for routing optimization (or path planning) in indirect satellite communications through multi-hop connectivity, which is a key challenge in massive \ac{leo} constellations. Overall, position information can enhance \ac{leo} satellite communications at both the link and network levels, and its role is particularly important in \ac{leo} systems because of their fast satellite motion, large coverage areas, and time-varying network topology.

\subsection{The Role of Communication in Enhancing Positioning}
Communication can also enhance \ac{leo}-based positioning by providing useful infrastructure, network coordination, and side information. This subsection examines how communication resources and communication-side information can support positioning.

\subsubsection{Infrastructure Perspective}
Broadband \ac{leo} communication systems often use high-frequency and directional signals to support high-throughput services. At higher frequencies, the shorter wavelength allows smaller antenna elements and inter-element spacing, making dense phased arrays practical within a limited satellite payload. Such arrays provide higher beamforming gain to improve the link budget. Although these infrastructures are primarily designed for communication, they can also provide useful geometric observables (e.g., \ac{aod} and \ac{aoa}) when broadband \ac{leo} signals are exploited for opportunistic positioning and auxiliary \ac{pnt} applications~\cite{10279786}. 
However, in multi-beam phased-array payloads, beam steering may introduce differential group delay and antenna-response mismatch across beams~\cite{delos2020phased, pohlmann2022simultaneous}, while Doppler or timing pre-compensation may further affect the apparent positioning observables. In addition, simultaneous reception from multiple satellites generally requires appropriate frequency separation to mitigate cross-satellite interference. 
Unlike \ac{gnss} signals using \ac{cdma}, which can support code-domain separation over overlapping time-frequency resources, many communication-oriented LEO systems employ \ac{ofdm} waveforms and generally do not support uncoordinated simultaneous transmissions over the same time-frequency resources. 
Therefore, the positioning gains offered by antenna arrays depend on accurate hardware and geometric calibration, as well as effective management of multi-source signal reception.

In \ac{5g}/\ac{6g} systems, the extensive terrestrial communication infrastructure can be leveraged and integrated with satellite-based positioning systems to enhance localization capabilities. Traditional \ac{gnss} applications already demonstrate the use of ground stations to assist high-precision positioning, as seen in \ac{rtk} positioning systems~\cite{Zheng20235G}. In emerging \ac{leo} satellite systems, theoretical studies have also shown that integrating ground-based technologies, such as \acp{ris}, can potentially improve positioning and tracking accuracy, particularly in complex suburban and urban environments~\cite{Zheng2024LEO}.

\subsubsection{Network Coordination Perspective}
In addition to the advantages provided by new communication infrastructures, efficient inter-satellite and satellite-to-ground communication can support positioning by enabling cooperation across the network. In particular, such links allow distributed positioning anchors, such as satellites or \acp{bs}, to exchange timing, ephemeris, calibration, and measurement information, thereby supporting networked positioning and joint signal processing~\cite{rs11182117,komodromos2025network}. For example, decentralized \ac{leo}-\ac{gnss} networks have been shown to approach centralized orbit/time-estimation performance with a reduced communication and processing burden~\cite{liu2025leo}. However, realizing such cooperation requires efficient protocols under limited \ac{isl} capacity.

\subsubsection{Other Benefits}
LEO communication can support clock synchronization by enabling the exchange of timing information between satellites and ground-based atomic clocks, thereby improving timing accuracy for positioning estimation~\cite{pan2018time}. Additionally, \ac{leo} communication facilitates fault-tolerant network operation by supporting vulnerable-node identification, control switching, and data rerouting in the event of satellite or ground station failures~\cite{ZHANG2025110699}. Such network resilience is important for maintaining continuous positioning services. Furthermore, the two-way communication capability of \ac{leo} systems enhances dynamic and responsive positioning services, which are difficult to realize in one-way systems such as conventional \ac{gnss}.

\subsection{Illustrative Case Study: Communication-Assisted Acquisition and Availability}
\begin{table*}[t]
\centering
\caption{Simulation Parameters for the Communication-Assisted LEO-PNT Case Study}
\label{table_4}
\begin{tabular}{@{}p{0.6\linewidth}p{0.4\linewidth}@{}}
\toprule
Parameter & Value \\
\midrule
\multicolumn{2}{@{}l}{\textit{--- Scenario and Geometry ---}} \\
Orbit altitude $h$ & \unit[550]{km} \\
Minimum elevation angle $\theta_\mathsf{min}$ & $40^\circ$ \\
Carrier frequency & \unit[2.2]{GHz} \\
Geometrically visible PNT satellites & $6$ \\
User test position (latitude, longitude, altitude) 
& $(45^\circ,\,0^\circ,\,\unit[0]{m})$ \\
Monte Carlo trials & $7000$ \\
\addlinespace
\multicolumn{2}{@{}l}{\textit{--- PNT Signal Model ---}} \\
PNT chip rate & \unit[4.092]{Mcps} \\
PN code period & $4092$ chips \\
PN code period duration & \unit[1]{ms} \\
PN periods per symbol & $10$ \\
Ranging integration time $T_{\mathsf{int}}$ & \unit[10]{ms} \\
Pseudorange noise standard deviation $\sigma_\mathsf{p}$ & \unit[2.85]{m} \\
\bottomrule
\end{tabular}
\end{table*}

To provide a numerical illustration of the preceding discussion, this subsection presents a representative case study on communication-assisted \ac{leo}-\ac{pnt} acquisition. We consider a downlink integrated communication and \ac{leo}-\ac{pnt} scenario based on the architecture proposed in~\cite{de2025integrated}, where communication-side information is used to assist positioning signal acquisition and provide auxiliary information for positioning. Specifically, the \ac{ue} aims to acquire the \ac{pnt} signals transmitted by geometrically visible satellites, while the communication link provides assistance information to reduce the acquisition search space. The simulation parameters are summarized in Table~\ref{table_4}, and the \ac{leo} signal measurement model follows~\cite[Eq.~2]{singh2022opportunistic}.

Building on this measurement model, the receiver first performs signal acquisition by searching over possible delay, Doppler, and code-phase hypotheses to identify the correct \ac{leo}-\ac{pnt} signals. In the considered communication-assisted setting, communication-side information narrows the uncertainty region by providing Doppler priors, and the remaining search space is represented by \(N_{\mathsf{res}}\) residual search bins. The receiver can test a finite number of these bins in parallel, referred to as the \textit{parallel search bins}.

After acquisition, positioning is performed using pseudorange measurements from the successfully acquired satellite signals. For satellite \(s\), the pseudorange measurement at \ac{ue} \(k\) is modeled as
 \begin{equation}
     \rho_{s,k} = \| {\bf p}_{\mathsf{u},k} -{\bf p}_s\|_2 + b_{\mathsf{u},k} + \delta_s + n_{s,k},
 \end{equation}
 where \({\bf p}_{\mathsf{u},k}\) and \({\bf p}_s\) are the positions of the \(k\)-th \ac{ue} and the \(s\)-th satellite, respectively; \(b_{\mathsf{u},k}\) is the receiver clock-bias term expressed in meters, \(\delta_s\) represents the residual satellite clock and ephemeris error term, and \(n_{s,k} \sim \mathcal{N}(0,\sigma_\mathsf{p}^2)\) is the ranging noise. The \ac{ue} position and receiver clock bias are then estimated from the acquired satellite measurements. A position fix is considered valid only if at least four satellites are successfully acquired, and the resulting \ac{pdop} is below the prescribed threshold.
 
\begin{figure}[ht]
    \centering
    \includegraphics[width=0.8\linewidth]{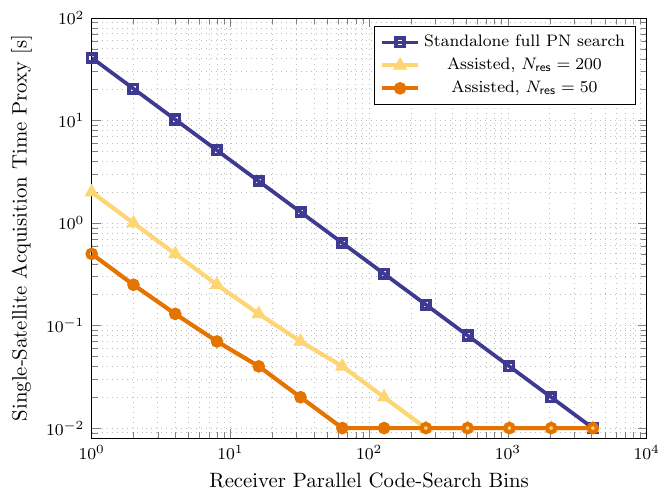}
    \caption{Single-satellite acquisition time proxy versus the number of receiver parallel code search bins. The blue solid curve with square markers represents the standalone full-PN-code search case. The yellow solid curve with triangle markers represents the communication-assisted case with a residual search window \(N_{\mathsf {res}}=200\), and the orange solid curve with circle markers represents the communication-assisted case with \(N_{\mathsf {res}}=50\). Here, \(N_{\mathsf {res}}\) denotes the number of residual code-search bins after communication-side assistance.}
\label{fig:case2_ttff_proxy}
\end{figure}

\begin{figure}[t]
    \centering
    \includegraphics[width=0.8\linewidth]{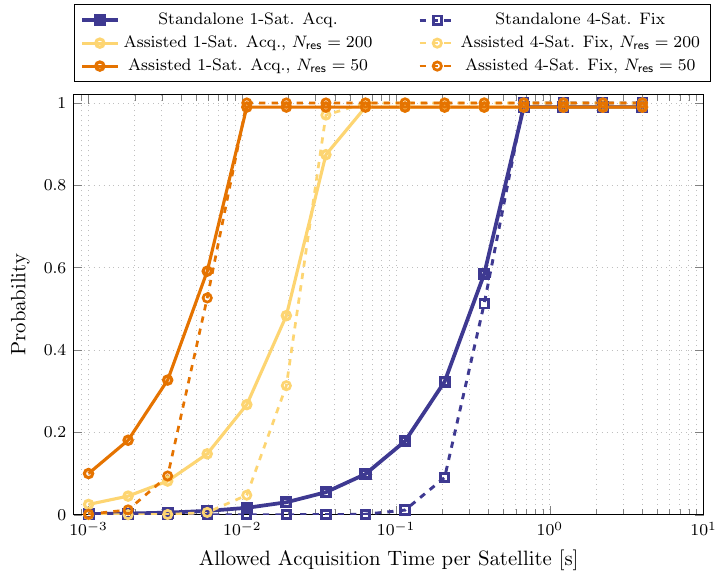}
    \caption{TTFF-limited positioning availability versus the allowed acquisition time per satellite. The blue curves represent the standalone case, the yellow curves represent the communication-assisted case with a residual search window \(N_{\mathsf {res}}=200\), and the orange curves represent the communication-assisted case with \(N_{\mathsf {res}}=50\). The solid curves represent the probability of single-satellite acquisition, while the dashed curves represent the probability of achieving a valid four-satellite position fix. Here, \(N_{\mathsf {res}}\) denotes the number of residual code-search bins after communication-side assistance. A successful position fix is defined as acquiring at least four satellites within the acquisition time constraint.}
\label{fig:case2_positioning_availability}
\end{figure}

Figure~\ref{fig:case2_ttff_proxy} shows the single-satellite proxy acquisition time with and without communication-side assistance. In the standalone case, the \ac{ue} searches the full pseudo-noise (PN) code space; therefore, the acquisition time remains high unless a very large number of parallel search bins are available. By contrast, with the assistance of communication signals, the residual search window is substantially reduced, and the corresponding acquisition time proxy is therefore significantly lower.
Figure~\ref{fig:case2_positioning_availability} shows how the reduction in \ac{ttff} improves positioning availability. The solid curves denote the single-satellite acquisition probability, while the dashed curves denote the probability of acquiring at least four satellites. Compared with the standalone case, the communication-assisted cases achieve a higher acquisition probability under the same allowed acquisition time. This improvement becomes more pronounced when the residual search window is smaller, as illustrated by the stronger gain in the \(N_{\mathsf{res}}=50\) case. Moreover, this benefit is not limited to single-satellite acquisition but translates directly into the multi-satellite availability required for positioning. Under the same time constraint, the probability of acquiring at least four satellites is significantly higher in the communication-assisted cases than in the standalone case.

\begin{table}[ht]
\centering
\caption{Results for Communication-Assisted LEO-PNT Positioning}
\label{table_5}
\begin{tabular}{lcc}
\toprule
Case & 95\% horizontal error (m) & Valid fixes (\%) \\
\midrule
Standalone, acquisition-limited & 82.99 & 83.54 \\
COM acquisition prior \((N_{\mathsf{res}}=50)\) & 42.28 & 99.70 \\
COM acquisition prior + ancillary navigation \((N_{\mathsf{res}}=50)\) & 30.35 & 99.66 \\
\bottomrule
\end{tabular}
\end{table}

Table~\ref{table_5} quantifies the positioning gains with the help of communication. Compared with the standalone acquisition-limited case, introducing the communication-assisted acquisition prior reduces the 95th-percentile horizontal positioning error from 82.99~m to 42.28~m, while increasing the valid-fix rate from 83.54\% to 99.70\%. When ancillary navigation information, i.e., side information that helps reduce residual satellite clock and ephemeris bias, is further introduced, the 95th-percentile horizontal error is further reduced to approximately 30~m, while the valid-fix rate remains nearly unchanged at about 99.7\%.

Overall, in this case study, these results illustrate that communication can support LEO positioning by improving acquisition efficiency, valid-fix availability, and final positioning accuracy.

\section{Open Problems}
Despite the promising potential of \ac{leo}-based \ac{ipac}, several practical and theoretical challenges remain before such systems can be widely deployed. This section summarizes key open problems for future \ac{leo}-based \ac{ipac} system design.

\subsection{Trade-Offs in Integrated Communication and Positioning}
In \ac{leo}-based \ac{ipac} systems, communication and positioning are not two independent functions simply deployed on the same satellite platform. Instead, they need to share common system resources, including spectrum, antenna apertures, beam patterns, signal structures, and network coordination mechanisms. Therefore, a central challenge in \ac{leo}-based \ac{ipac} is to balance the requirements of communication and positioning through joint system design. This subsection discusses three representative trade-offs related to terminal and antenna architecture, beam operation and signal pre-compensation, and resource and spectrum allocation.
\subsubsection{Terminal and Antenna Architecture}

Terminal and antenna architecture introduce a fundamental trade-off between communication performance and positioning availability. In mobile-satellite and \ac{d2d}/\ac{ntn} systems operating in the L/S-band or other lower-frequency cellular spectrum, user terminals are typically handheld devices or \ac{iot} devices equipped with low-gain or quasi-omnidirectional antennas~\cite{de2025integrated}. Such architectures are favorable for positioning because they allow a receiver to observe signals over a wider angular range and are therefore more suitable for simultaneous multi-satellite reception. However, these systems usually provide smaller bandwidth than Ku/Ka-band broadband systems, which limits delay resolution. 

By contrast, broadband \ac{leo} systems operating mainly in Ku/Ka bands generally employ directional or electronically steerable antennas to provide high link gain, spatial isolation, and throughput. The larger bandwidth of these systems is also attractive for positioning since it can improve delay-domain resolution. However, their directional architecture may restrict the reception geometry and reduce the opportunity for simultaneous reception from multiple satellites. In addition, accurate positioning based on such systems requires careful calibration of satellite- and terminal-side group delay, phase response, and antenna-dependent biases. 
Therefore, terminal and antenna design in \ac{leo}-based \ac{ipac} must balance link budget, bandwidth, angular visibility, multipath robustness, and calibration complexity.

\subsubsection{Beam Operation and Observable Modeling}
Beam operation introduces another important trade-off between communication service continuity and positioning observable interpretability. As discussed earlier, in \ac{leo} communication systems, Doppler and timing pre-compensation are often performed with respect to the beam center or a reference point within the service area. In addition, beam footprints may be steered in an Earth-fixed or moving-cell manner to reduce handover frequency and simplify network operation~\cite{3GPP38821}. Together with Doppler and timing pre-compensation, these mechanisms are introduced to mitigate the large time and frequency offsets caused by \ac{leo} satellite motion and propagation delay, thereby reducing the synchronization burden on the \ac{ue}. However, these operations can introduce beam-dependent corrections into the received signal. As a result, when communication signals are exploited for opportunistic \ac{pnt}, the extracted delay and Doppler observables may no longer correspond directly to the satellite-user geometry unless the beam-dependent pre-compensation is explicitly modeled and available.

\subsubsection{Resource Sharing and Spectrum Planning}

Resource sharing and spectrum planning introduce another trade-off between communication efficiency and positioning availability. In LEO communication systems, a UE is typically associated with a serving satellite, cell, or beam for data transmission, while spatial, frequency-domain, or time-domain resource separation is used to mitigate inter-beam or inter-satellite interference and improve resource utilization. Such communication-oriented resource allocation may limit the simultaneous availability of signals from multiple satellites, which is unfavorable for positioning because measurements from multiple satellites are generally required for a position fix. This issue is particularly relevant to D2D/NTN systems, where simultaneous reception from multiple satellites by low-complexity terminals may require appropriate frequency separation or frequency coloring to avoid cross-satellite interference. Therefore, positioning-aware resource allocation and multi-satellite signal scheduling are needed to jointly support communication efficiency and positioning availability~\cite{de2025integrated,de2026beyond1,de2026beyond}.

\subsection{GNSS Air-Interface Adaptation to LEO}

Another open issue is whether, and to what extent, existing GNSS or standardized communication air interfaces can be adapted to LEO-PNT systems. In practice, most commercial LEO constellations operate in non-GNSS bands (see Table~\ref{table_1}), and their communication waveforms were not originally designed for navigation or ranging. On the one hand, GNSS L1/L5 air interfaces cannot be simply repurposed for broadband communication services, since these bands are narrowly allocated and tightly regulated for navigation use. Even dedicated LEO-PNT systems operating near GNSS L-band frequencies, such as Xona Pulsar, must carefully control spectrum usage and signal design to coexist with legacy GNSS services. On the other hand, reusing terrestrial communication waveforms, such as \ac{5g} \ac{nr} \ac{ntn} signals or \ac{prs}-like structures, is also challenging in LEO environments. As a result, GNSS-compatible and communication-native air interfaces are unlikely to converge into a single straightforward solution. Determining how to balance waveform compatibility, receiver complexity, Doppler resilience, spectrum coexistence, and positioning performance therefore remains an important open problem for future LEO-PNT and integrated communication-positioning system design.

    \subsection{Impact of Orbit Errors}
    The orbit information of LEO satellites is often inaccurate due to several factors, such as perturbing forces, atmospheric drag, and solar radiation pressure. Moreover, the process of parameterizing orbital data into two-line element (TLE) files introduces errors due to the constraints of the limited format~\cite{acciarini2025closing}. Combined with the use of simplified models for orbit propagation, these factors introduce systematic orbital errors that degrade satellite state prediction accuracy~\cite{ma2025positioning}. This not only impairs positioning performance but also degrades communication quality by causing beam-pointing bias and thereby reducing link reliability and achievable rates~\cite{11014543}. As shown in~\cite{ma2025positioning}, even with more observations or increased transmit power, the bias induced by orbital mismatch cannot be removed. For \ac{ipac} implementation, it is critical (i) to jointly consider ephemeris freshness and uncertainty, and (ii) to develop orbit-aware adaptive approaches that adjust beam pointing and Doppler pre-compensation as ephemeris uncertainty evolves.

    Several approaches have been proposed to mitigate this critical issue. Observation-driven methods exploit real-time measurements, such as Doppler or differential observations, to refine orbit- or clock-related information~\cite{wang2023doppler,xie2025differential,hasan2024double}. In contrast, ephemeris-driven methods rely on historical orbital information to characterize and compensate for orbital errors~\cite{gleba2025impact,caldas2024precise}. Nevertheless, developing more robust, timely, and scalable correction methods remains an important direction for future \ac{ipac} systems.

\subsection{Inter- and Intra-Network Cooperation}

    In \ac{leo}-based \ac{ipac} systems, cooperation between the \ac{tn} and the \ac{ntn} is important for combining their complementary strengths in coverage, integrity, and resilience~\cite{saleh2025integrated}. On the ground side, dense terrestrial infrastructures can provide high-resolution angle, delay, and Doppler measurements; on the space side, \ac{leo} satellites provide wide-area coverage, frequent visibility opportunities, and geometry diversity. By jointly exploiting these complementary resources, inter-network cooperation can support more robust communication and positioning services across large geographic regions.

 Beyond \ac{tn}-\ac{ntn} cooperation, intra-network cooperation within the satellite segment is also essential for LEO-based IPAC. Inter-satellite links and satellite-to-ground links can enable the exchange of timing, ephemeris, calibration, and measurement information among satellites, gateways, and ground stations. Such information-level cooperation can support distributed positioning, orbit and clock correction dissemination, network-level resource coordination, and joint estimation~\cite{rs11182117,liu2025leo}. These cooperative functions require efficient protocols and compact signaling to operate under limited \ac{isl} capacity, link intermittency, and time-varying satellite topology, making low-overhead cooperation an important open problem.

    \subsection{Satellite Handover Management}
    Satellite handover refers to the process of switching the serving satellite or active satellite set for communication and positioning as satellites move across the coverage area of a ground station or the signal reception range of a \ac{ue}. Due to their rapid movement relative to the Earth's surface, \ac{leo} satellites require frequent handovers to ensure continuous communication and accurate positioning. This regular switching intensifies the complexity of network resource management and challenges the maintenance of uninterrupted service continuity. Managing handovers within a constellation of potentially hundreds or thousands of satellites demands advanced network management systems and sophisticated algorithms to coordinate traffic and enhance network performance.

    \subsection{Positioning Privacy and Communication Security}
    \Ac{leo} \Ac{ipac} systems also introduce substantial privacy and security challenges. For example, enabling position-aware beamforming requires access to \ac{ue} position information. Sharing such information may increase the risk of position privacy leakage, as the broadcast nature of wireless channels exposes position information to unauthorized nodes. Additionally, the wide coverage of \ac{leo} satellite networks makes them highly vulnerable to eavesdropping, especially from unauthorized entities near the UEs. Conventional secure beamforming techniques, which focus on directing transmit power to legitimate nodes while suppressing it towards illegitimate ones, may be insufficient in \ac{leo} scenarios. The broader beam footprint on the ground, compared with that of terrestrial systems, complicates precise and secure beamforming, thus presenting unique challenges for safeguarding \ac{leo} \ac{ipac} systems.

\section{Conclusion}
This paper provides a contemporary overview of IPAC systems enabled by LEO satellites. By reviewing representative LEO constellations, positioning mechanisms, communication characteristics, and open challenges, we show that positioning and communication are increasingly interconnected in future LEO networks. The representative case study further illustrates that communication-side assistance can reduce the acquisition burden of LEO-PNT signals and improve positioning availability and accuracy. These observations suggest that future LEO-based IPAC systems should be designed by jointly considering signal design, beam operation, ephemeris and timing information, and resource allocation.

\backmatter

\section*{Acknowledgments}
Figure~\ref{fig:workflow} was created by Ana Runte, a Scientific Illustrator at KAUST. This publication is based upon work supported by King Abdullah University of Science and Technology (KAUST) under Award No. ORFS-CRG12-2024-6478. The authors acknowledge this funding support.

\section*{Author contributions}
J.M. performed the simulations, prepared the figures, and drafted the manuscript. P.Z. developed the main idea, contributed to the simulations, co-drafted and revised the manuscript. X.L. contributed to manuscript writing and revisions. Y.Z. contributed to manuscript writing and revisions. T.Y.A.-N. supervised the project and provided critical revisions. All authors read and approved the final manuscript.

\section*{Competing interests}
T.Y. Al-Naffouri serves as an editor for this journal and is a co-author of this manuscript. To avoid any potential conflict of interest, he had no involvement in the editorial evaluation, peer-review process, or final decision for this submission. The remaining authors declare no competing interests.

\section*{Ethics approval and consent to participate} Not applicable.

\section*{Consent for publication} Not applicable.

\section*{Data availability} Not applicable.

\section*{Materials availability} Not applicable.

\section*{Code availability} Not applicable.








\section*{Figure captions}

\textbf{Figure~\ref{fig:workflow}. Structure of LEO satellite systems, divided into the space, user, and ground segments. The space segment of the system provides service and feeder links with various payload types, including regenerative and transparent payloads. The user segment encompasses ground users, infrastructures, and building environments. The ground segment consists of gateways, base stations, data centers, core networks, and other related components. }

\textbf{Figure~\ref{fig:case2_ttff_proxy}. Single-satellite acquisition time proxy versus the number of receiver parallel code search bins. The blue solid curve with square markers represents the standalone full-PN-code search case. The yellow solid curve with triangle markers represents the communication-assisted case with a residual search window \(N_{\mathsf {res}}=200\), and the orange solid curve with circle markers represents the communication-assisted case with \(N_{\mathsf {res}}=50\). Here, \(N_{\mathsf {res}}\) denotes the number of residual code-search bins after communication-side assistance. }

\textbf{Figure~\ref{fig:case2_positioning_availability}. TTFF-limited positioning availability versus the allowed acquisition time per satellite. The blue curves represent the standalone case, the yellow curves represent the communication-assisted case with a residual search window \(N_{\mathsf {res}}=200\), and the orange curves represent the communication-assisted case with \(N_{\mathsf {res}}=50\). The solid curves represent the probability of single-satellite acquisition, while the dashed curves represent the probability of achieving a valid four-satellite position fix. Here, \(N_{\mathsf {res}}\) denotes the number of residual code-search bins after communication-side assistance. A successful position fix is defined as acquiring at least four satellites within the acquisition time constraint.}
\end{document}